\documentclass[aps,prl,twocolumn,superscriptaddress,floatfix]{revtex4-2}
\usepackage{color,graphicx}
\usepackage{amssymb,amsmath,bm}

\begin{document}
    \author{Shin Miyahara}
  \affiliation{Department of Applied Physics, Fukuoka University, Fukuoka 814-0180, Japan}

  \title{
    Theory of absorption in a frustrated spin gap system
    ${\rm SrCu_2(BO_3)_2}$ 
  }
  \date{\today}

  \begin{abstract}
    We achieve a comprehensive understanding of the magnetic excitations observed by electric spin resonance and far-infrared spectroscopy in a frustrated spin gap system ${\rm SrCu_2(BO_3)_2}$ by considering the effects of magnetoelectric couplings and Dzyaloshinskii-Moriya interactions in the Shastry-Sutherland model. The transitions from the dimer singlet ground state to the triplon and the bound states of two triplons are electroactive through the magnetoelectric couplings, even in the Shastry-Sutherland Heisenberg model. The results indicate that an electro-triplon and a novel electroactive magnetic excitation, as with an electromagnon in the multiferroics, can be realized in the conventional spin-gapped singlet and anomalous spin-liquid states. Clarifying these electroactive magnetic excitations in various spin systems will help analyze a broad range of quantum magnets, {\it e.g.}, quantum spin liquids and spin nematics.
  \end{abstract}

  \maketitle
 
  The quasi-two-dimensional magnet ${\rm SrCu_2(BO_3)_2}$~\cite{smith91,kageyama99}  is a realization of the Shastry-Sutherland model, of which the ground state is precisely described as a direct product of dimer singlet states~\cite{shastry81}, and  shows anomalous behaviors, {\it e.g.},  spin-singlet ground states~\cite{kageyama99,miyahara99},  almost localized triplon excitation~\cite{miyahara99,kageyama00c,romhanyi15,suetsugu22},  magnetization plateaus~\cite{onizuka00,kodama02,matsuda13,corboz14} and  pressure-induced quantum phase transitions~\cite{waki07,zayed17,sakurai18,guo20,jimenez21,shi22}. These magnetic behaviors have been studied  enthusiastically from experimental and theoretical viewpoints as a typical example of two-dimensional frustrated magnets~\cite{miyahara03}.

  Spin excitation spectra in ${\rm SrCu_2(BO_3)_2}$ have been investigated by electron spin resonance (ESR)~\cite{nojiri99,nojiri03,sakurai18},  far-infrared (FIR) spectroscopy~\cite{room00,room04} and Raman scatterings~\cite{lemmens00b,wulferding21}. The resonance due to a triplon and bound states of two triplons~\cite{momoi00,momoi00b,fukumoto00b,totsuka01}  have been observed.
  However, these magnetic transitions are forbidden in a singlet ground state of a Heisenberg model. Although the several possible mechanisms of the triplon transition,  {\it e.g.}, the effects of DM interactions,  the spin-phonon interactions, and/or magnetoelectric couplings were proposed~\cite{sakai00,cepas01,nojiri03,room04,kimura18}, a comprehensive understanding of the  mechanism and the selection rule in absorption in ${\rm SrCu_2(BO_3)_2}$ has not yet been established.

  In this Letter, we consider  the effects of the couplings between electric polarization and  spin moments~\cite{tokura06,eerenstein06,kimura07,nagaosa08,arima11,nagaosa12,tokura14}. We show that the electric field of light can excite a triplon excitation and bound states of two triplons in the Heisenberg model on the Shastry-Sutherland lattice, as with an electromagnon~\cite{katsura07,aguilar09} and two-magnon excitations~\cite{moriya66,moriya68,akaki17} in ordered magnets, and a spin gap in dimer systems~\cite{kimura15,kimura18}. By considering the effects of the magnetoelectric couplings and the anisotropic spin interactions in ${\rm SrCu_2(BO_3)_2}$, we can explain the resonance peak positions and the selection rules observed in ${\rm SrCu_2(BO_3)_2}$.

\begin{figure}
    \begin{center}
      \includegraphics[width=1.0\columnwidth,bb= 0 0 2427 1465]{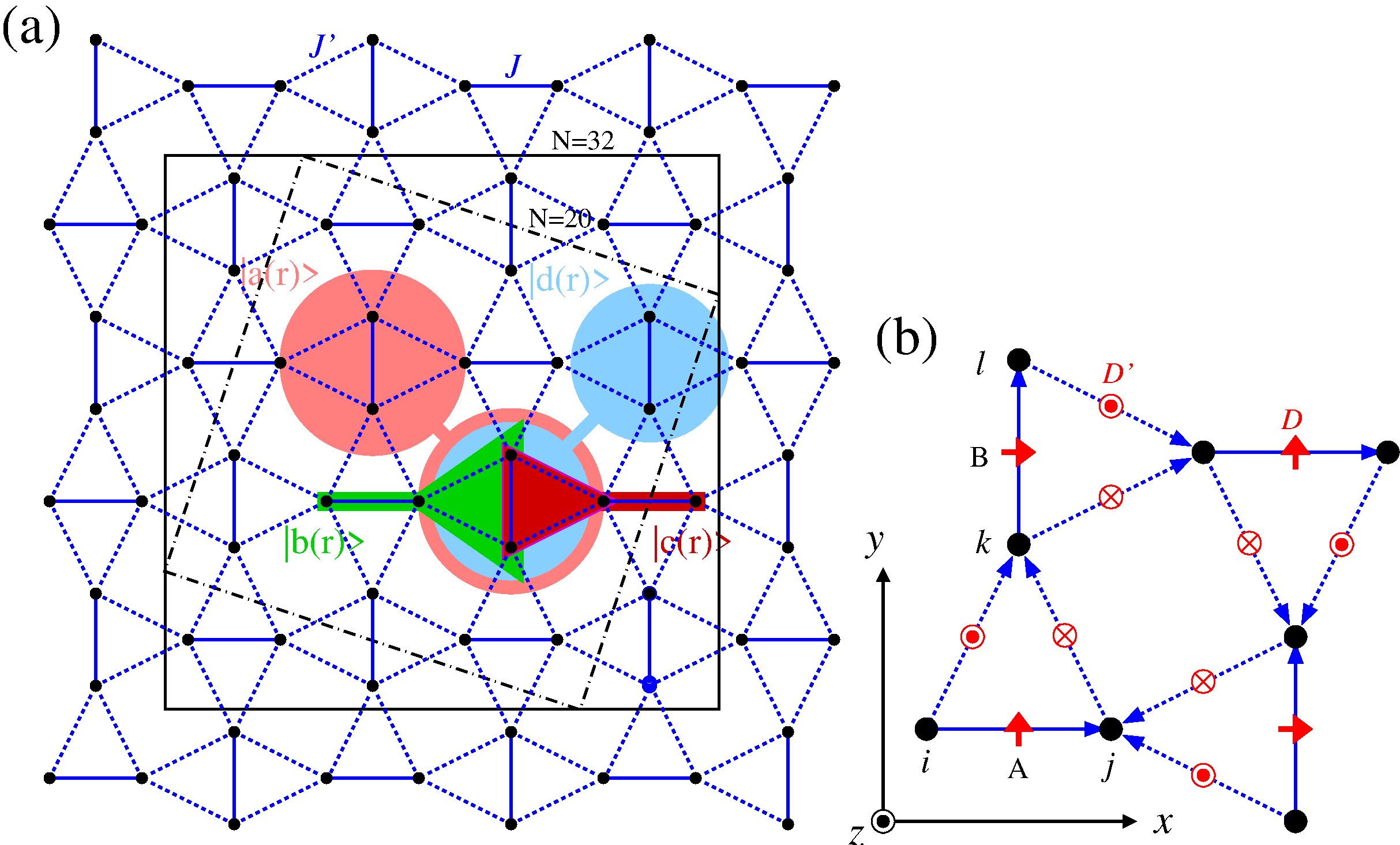}
    \end{center}
    \caption{(Color online)
      (a) Shastry-Sutherland model. The four configurations of two-triplon bound states $|a(\bm r)\rangle$ $|b(\bm r)\rangle$, $|c(\bm r)\rangle$, $|d(\bm r)\rangle$, and 20-site and 32-site cluster. 
      (b) The blue (red) arrows define the bond directions (DM vectors).}
    \label{fig:lattice}
  \end{figure}

  The magnetic properties of ${\rm SrCu_2(BO_3)_2}$ under external magnetic field ${\bm B}^{\rm ex}$ can be explained by a spin $S=1/2$ Hamiltonian on the Shastry-Sutherland lattice [Fig.~\ref{fig:lattice} (a)],
  \begin{align}
    H_{\rm S}  = & J \sum_{\rm n.n.} {\bm S}_i \cdot {\bm S}_j
    +  J^\prime \sum_{\rm n.n.n.} {\bm S}_i \cdot {\bm S}_j
    - \mu_{\rm B} \,\, \sum_i  {\bm B}^{\rm ex} \cdot \hat{g} {\bm S}_i,
  \end{align}
  where $J$ and $J^\prime$ are the intradimer and interdimer interactions. $\mu_{\rm B}$ is the Bohr magneton, and $\hat{g}$ is a $g$-tensor. The diagonal components are estimated to be $g_x = g_y = 2.05$, and $g_z = 2.28$~\cite{nojiri99}. We neglect the off-diagonal components of $\hat{g}$ for simplicity.

  The couplings between electric polarization and spin moments can be introduced through spin-dependent electronic polarization~\cite{moriya68,katsura05,kaplan11}. We consider the electric polarizations couple to a symmetric spin pair ${\bf P}^{\rm S}$ on the $J^\prime$ bond and an antisymmetric spin pair ${\bf P}^{\rm AS}$ on the $J$ bond, which are
  \begin{align}
    {\bf P}^{\rm S} & = \sum_{\rm n.n.n.} \Pi \, {\bm e}_{ij} \, {\bm S}_i \cdot {\bm S}_j,
    & \label{eq:P_S} & \\
    {\bf P}^{\rm AS} & = \sum_{\rm n.n.}
    d \, {\bm e}_{ij} \times \left( {\bm S}_i \times {\bm S}_j \right).
    \label{eq:P_AS} &
  \end{align}
  The direction of the bond ${\bf e}_{ij} (|{\bf e}_{ij}|=1)$ is defined in Fig.~\ref{fig:lattice} (b). We assume that the direction of ${\bm \Pi}_{ij}$ is parallel to that of the next nearest neighbor (n.n.n.) bond ${\bf e}_{ij}$. We assume that the crystal structure of ${\rm SrCu_2(BO_3)_2}$ is the tetragonal ${\rm I\bar{4}2m}$ space group~\cite{smith91,sparta01} for simplicity. Furthermore, we neglect the ${\bf P} \propto {\bf S}_i \times  {\bf S}_j$ on the n.n.n. bonds, and the symmetric anisotropic term of ${\bf P}$ like $P \propto S_i^z S_j^x + S_i^x S_j^z$.

  We calculate the dynamical electric susceptibility $\chi^{{\rm ee} \alpha}_{\beta \beta} (\omega)$ to clarify the electroactive excitation processes in the Shasry-Sutherland-Heisenberg model.
  \begin{align}
    \chi^{{\rm ee} \alpha}_{\beta \beta} (\omega) & = 
    \frac{1}{\hbar N V \epsilon_0}
    \sum_{n} \frac{\langle 0 |  P_\beta^\alpha | n \rangle
      \langle n |  P^\alpha_\beta | 0 \rangle}
	{\omega_{n0} - \omega -  i\delta}, &
    \label{eq:chi-ee}
  \end{align}
  where $\alpha = {\rm S, \,\,\,AS}$ and $\beta = x, y, z$. $|0\rangle$ is a ground state with an eigenenergy $E_0$ and $|n\rangle$ is an $n$-th magnetic excitation state with an eigenenergy $E_n$. Here, $\hbar \omega_{n0} = E_n - E_0$, $V$ is a unit volume per spin, $N$ is the number of spins, and $\epsilon_0$ is the permittivity in vacuum.
  We fix that $\Pi=1$, $d=1$, $\hbar N V \epsilon_0 = 1$, and $\delta/J = 0.005$. $\chi^{{\rm ee} \alpha}_{\beta \beta} (\omega)$ is calculated on $20$- and $32$-site clusters using the Lanczos method~\cite{dagotto94,wietek18}.

\begin{figure}
    \begin{center}
      \includegraphics[width=1.0\columnwidth,bb= 0 0 640 480]{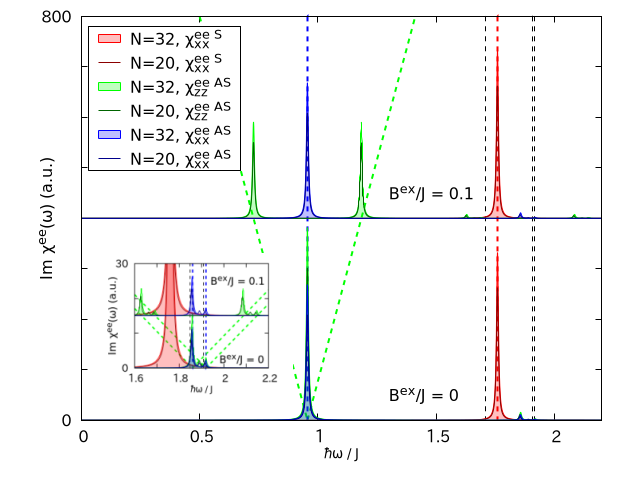}
    \end{center}
    \caption{(Color online)
      The imaginary parts of the dynamical electric susceptibility $\chi^{{\rm ee} \alpha}_{\beta \beta} (\omega)$ on the $20$- and $32$-site clusters for $J^\prime/J = 0.2$ under the external magnetic field $B^{\rm ex}_z/J = 0$ and $0.1$. The dotted lines show the spin gap and the $S^{\rm tot} = 0$ bound state energy obtained by third-order perturbation calculations. The red, blue, and green dotted lines, respectively, indicate the active modes for $E \perp z$ through ${\rm P}^{\rm S}$, for $E \perp z$ through ${\rm P}^{\rm AS}$, and for $E \| z$ through ${\rm P}^{\rm AS}$.
      (inset) The magnified view in the range $1.6 \leqq \hbar \omega/J  \leqq 2.2$. The blue and green dotted lines, respectively, indicate the active modes for $E \perp z$ through ${\rm P}^{\rm AS}$ and for $E \| z$ through ${\rm P}^{\rm AS}$ due to the $S^{\rm tot} = 1$ bound state excitations obtained by third-order perturbation calculations.
    }
    \label{fig:res_Jp0.2}
  \end{figure}

  The results for $J^\prime/J = 0.2$ under the external magnetic field $B^{\rm ex}_z/J = 0$ and $0.1$ are presented in Fig.~\ref{fig:res_Jp0.2}. ${\rm Im} \chi^{{\rm ee} {\rm AS}}_{x x} (\omega) [= {\rm Im} \chi^{{\rm ee} {\rm AS}}_{y y} (\omega)]$, ${\rm Im} \chi^{{\rm ee} {\rm AS}}_{z z} (\omega)$, and ${\rm Im} \chi^{{\rm ee} {\rm S}}_{x x} (\omega) [= {\rm Im} \chi^{{\rm ee} {\rm S}}_{y y} (\omega)]$ show resonance peaks. We perform the third-order perturbation to clarify the resonance origin and find the nonzero terms of $\langle n |  P^\alpha_\beta | 0 \rangle$ for the triplon and the bound states of two triplons. Note that the bound states are linear combinations of four two-triplon states $|a ({\bm r}) \rangle$, $|b ({\bm r}) \rangle$, $|c ({\bm r}) \rangle$, and  $|d ({\bm r}) \rangle$ in Fig.~\ref{fig:lattice} (a), with a total spin of $S^{\rm tot} = 0, 1$, and $2$~\cite{momoi00,momoi00b,fukumoto00b,totsuka01}.
  The results indicate that the $S^z = 0 (\pm 1)$ triplon and two of the four modes of the $S^{\rm tot}=1$ [$S^z = 0 (\pm 1)$] bound states are excited by ${\bm E}(\omega) \| z$ $(x,y)$ through the coupling ${\bm P}^{\rm AS}$ and one of the four modes of the $S^{\rm tot}=0$ bound states is excited by ${\bm E}(\omega) \| x,y$ through the coupling ${\bm P}^{\rm S}$. The perturbation calculation results correlate with the exact diagonalization (Fig.~\ref{fig:res_Jp0.2}). Fig.~\ref{fig:res_Jp0.2} shows that the resonance intensities of the triplon and $S^{\rm tot} = 0$ bound state excitations are much stronger than those of the  $S^{\rm tot} = 1$ bound state.

\begin{figure}
    \begin{center}
      \includegraphics[width=1.0\columnwidth,bb= 0 0 800 500]{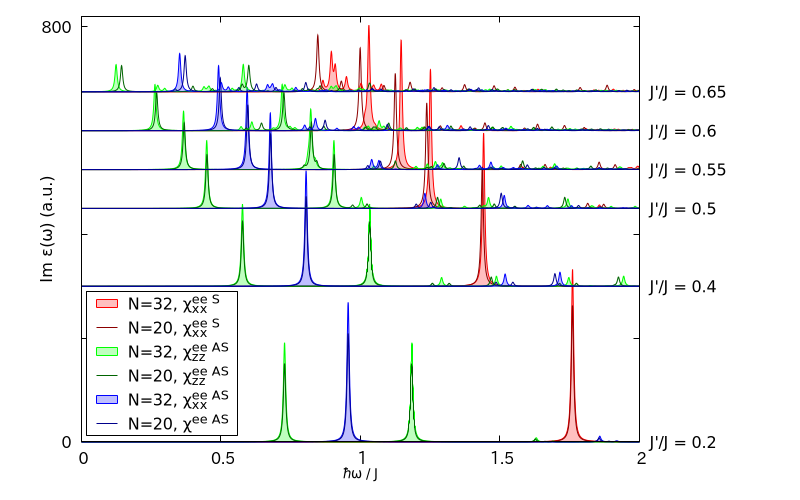}
    \end{center}
    \caption{(Color online)
      The imaginary parts of the dynamical electric susceptibility $\chi^{{\rm ee} \alpha}_{\beta \beta} (\omega)$ on the $20$- and $32$-site clusters when $J^\prime/J = 0.2, 0.4, 0.55, 0.6$, and $0.65$ under the external magnetic field $B^{\rm ex}_z/J = 0.1$.
    }
    \label{fig:res_Heisenberg}
  \end{figure}

  The results in the dimer singlet phase ($J^\prime/J = 0.2, 0.4, 0.55, 0.6$, and $0.65$) under the external magnetic field $B^{\rm ex}_z/J = 0.1$ are shown in Fig.~\ref{fig:res_Heisenberg} and the tendency of the resonance hardly depend on the parameter $J^\prime/J$. Thus, we conclude that the primary resonance peaks in the dimer singlet phase on the Shasry-Sutherland-Heisenberg model are the $S^z =0$ ($S^z = \pm 1$) triplon branch [${\bm E} \| x,y$ (${\bm E} \| z$)-active] and one of the $S^{\rm tot} = 0$ bound state branches (${\bm E} \| x,y$-active).

  To assign the peaks observed in the experiments~\cite{nojiri99,nojiri03,room00,room04}, the effects of the Dzyaloshinskii-Moriya (DM) interactions, which lift the degeneracy of the magnetic excitations~\cite{cepas01}, should be considered. The DM terms in ${\rm SrCu_2(BO_3)_2}$ are given by
  \begin{align} 
    H_{\rm D} = & D \left[ \sum_{\rm A} (S_i^z S_j^x - S_i^x S_j^z)
    + \sum_{\rm B} (S_k^y S_l^z - S_k^z S_l^y) \right] &  \nonumber \\
    & + D^\prime_{ij} \sum_{n.n.n.} (S_i^x S_j^y - S_i^y S_j^x), &
  \end{align}
  where $D$ and $D^\prime_{ij} (|D^\prime_{ij}| = D^\prime)$ are the intradimer and interdimer DM interactions, as shown in Fig.~\ref{fig:lattice} (b). The bond directions, $i \rightarrow j$ and $k \rightarrow l$, and the sign of $D^\prime_{ij}$ are also defined in Fig.~\ref{fig:lattice} (b). The parameter values for ${\rm SrCu_2(BO_3)_2}$ are fixed to be $J^\prime/J = 0.635$, $D/J = 0.034$, $D^\prime/J = 0.02$, and $J = 85$ K~\cite{kodama05}. We neglect the in-plane components of the interdimer DM interaction for simplicity.
  
  Since anisotropic interaction terms allow a transition from a singlet ground state to a triplon~\cite{sakai00}, we must consider the magnetic (M1 transitions) and electric processes (E1 transitions) in Eq.~(\ref{eq:chi-ee}). The magnetic processes can be read from the dynamical magnetic susceptibility $\chi^{\rm mm}_{\beta \beta} (\omega)$,
  \begin{align}
    \chi^{\rm mm}_{\beta \beta} (\omega) & = 
    \frac{\mu_0}{\hbar N V}
    \sum_{n} \frac{\langle 0 | M_\beta | n \rangle
      \langle n |  M_\beta | 0 \rangle}
	{\omega_{n0} - \omega -  i\delta}. &
	\label{eq:chi-mm}	
  \end{align}
  Here, the $\beta$ component of magnetization $M_\beta$ is $\displaystyle M_\beta = \sum g_\beta \mu_{\rm B} S_i^\beta$ and $\mu_0$ is the magnetic permeability in a vacuum. The other parameters are defined in Eq.~(\ref{eq:chi-ee}). We fix that $\mu_0 \mu_{\rm B}^2 / \hbar N V \epsilon_0 = 1$.

\begin{figure}
    \begin{center}
      \includegraphics[width=0.98\columnwidth,bb=0 0 640 480]{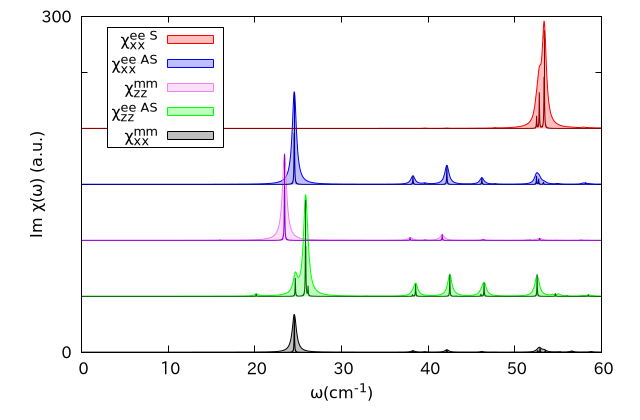}
    \end{center}
    \caption{(Color online)
      The imaginary parts of dynamical susceptibility for ${\bm B}^{\rm ex} = 0$ in the $N=20$ cluster. The parameters are fixed to be $J^\prime/J = 0.635$, $D/J = 0.034$, $D^\prime/J = 0.02$, and $J = 85$ K.}
    \label{fig:srcu2bo32_N20}
  \end{figure}

  The spectra of the imaginary parts of dynamical susceptibility calculated using the Lanczos method in the $N=20$ cluster for ${\rm SrCu_2(BO_3)_2}$ are shown in Fig.~\ref{fig:srcu2bo32_N20}.  Table \ref{tab:gap} summarizes the peak positions and selection rule corresponding to the spin gap ($S^{\rm tot}=1$) and $S^{\rm tot}=0$ bound state excitations in Fig.~\ref{fig:srcu2bo32_N20}.
  In the FIR spectroscopy, the strong resonances were observed at $\omega = 24.1 \, {\rm cm}^{-1}$ in ${\bf E} \| x$, $\omega = 25.5 \, {\rm cm}^{-1}$ in ${\bf E} \| z$, and $\omega = 52.2$ and $53.4  \, {\rm cm}^{-1}$ in ${\bf E} \| x$ through electroactive processes~\cite{room00,room04}. Note that the resonances at approximately  $\omega = 53 \,{\rm cm}^{-1}$ are much stronger than the other two and can be explained by the magnetoelectric couplings $\Pi \gg d$, where the spin-orbit interactions induce the $d$ but the $\Pi$ is caused only through exchange processes~\cite{moriya66,moriya68,katsura05,miyahara16}. As shown in Table \ref{tab:gap}, the experimental observations of the selection rules and excitation energies are consistent with the theoretical calculations in Fig.~\ref{fig:srcu2bo32_N20}, and the electroactive processes are dominant in ${\rm SrCu_2(BO_3)_2}$.

  \begin{table}
    \caption{The spin gap ($S^{\rm tot}=1$) and $S^{\rm tot}=0$ bound state excitation energy $\hbar \omega$ under a zero magnetic field ($B^{\rm ex} = 0$).
      $\omega$ (${\rm cm}^{-1}$) observed in FIR\cite{room04} and theory.
      The selection rules of ${\rm E1}$ due to the ${\rm P}^{\rm AS}$
      and ${\rm P}^{\rm S}$ and ${\rm M1}$ processes predicted in theory.}
    \label{tab:gap}
    \begin{tabular}{ccccccccc}
      \hline \hline
      & \hspace*{1cm} & \multicolumn{4}{c}{$S^{\rm tot} = 1$}
      & \hspace*{1cm} & \multicolumn{2}{c}{$S^{\rm tot} = 0$} \\ \hline
      FIR & & 22.7 & \multicolumn{2}{c}{24.1} & 25.5 & & 52.2 & 53.4\\
      theory & & 23.4 & 24.5 & 24.7 & 25.9 & & 52.8 & 53.4   \\
      ${\rm E1}$ (${\bm P}^{\rm AS}$) & & & $x,y$ & $z$ & $z$ & &  \\
      ${\rm E1}$ (${\bm P}^{\rm S}$) & & & & & & &  $x,y$ & $x,y$ \\
      ${\rm M1}$ (${\bm M}$) & & $z$ & $x,y$ & & & & & \\
      \hline  \hline
    \end{tabular}
  \end{table}
 
  \begin{figure}
    \begin{center}
      \includegraphics[width=0.48\columnwidth, bb= 0 0 4347 3531]{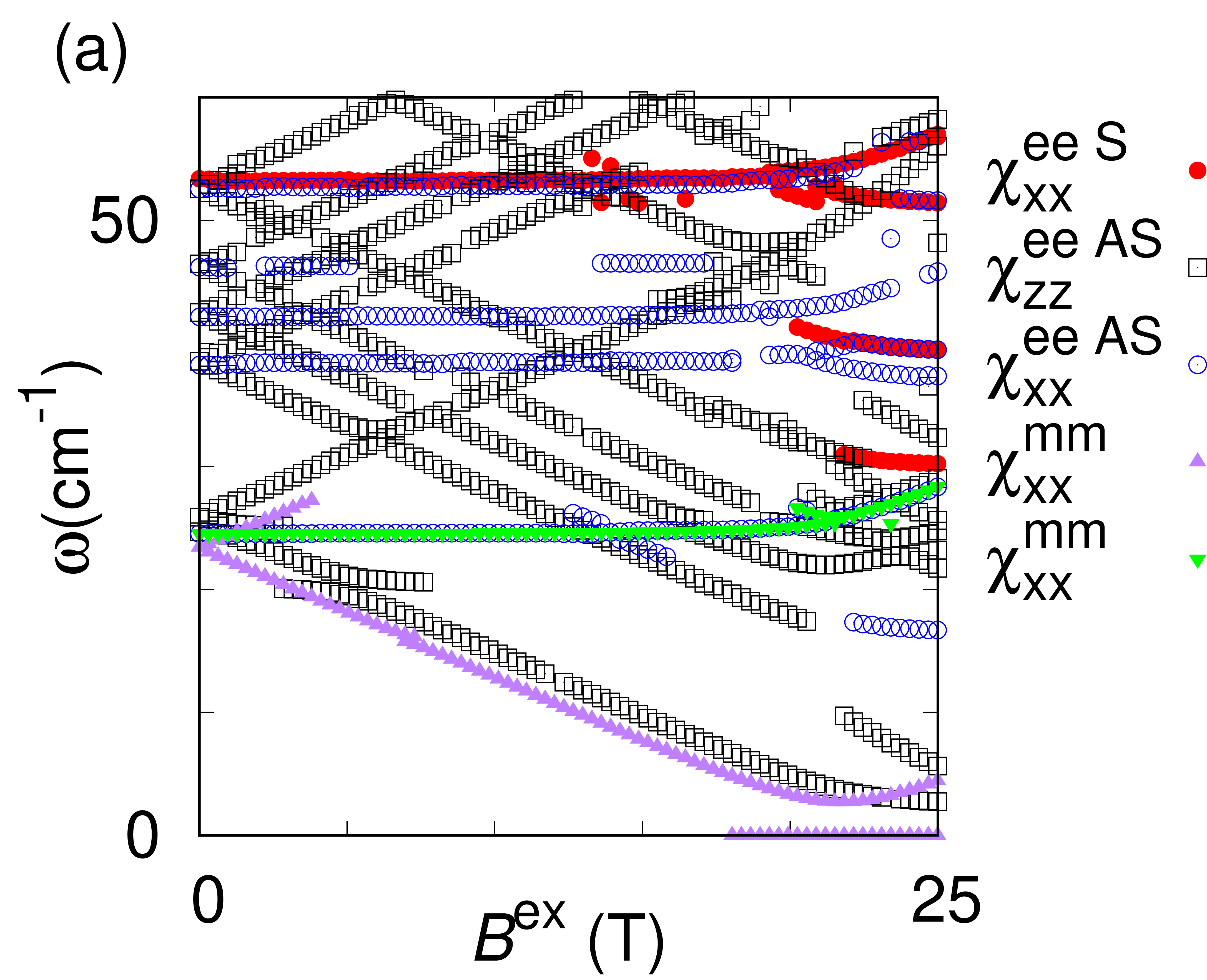}
      \includegraphics[width=0.48\columnwidth, bb= 0 0 4541 3929]{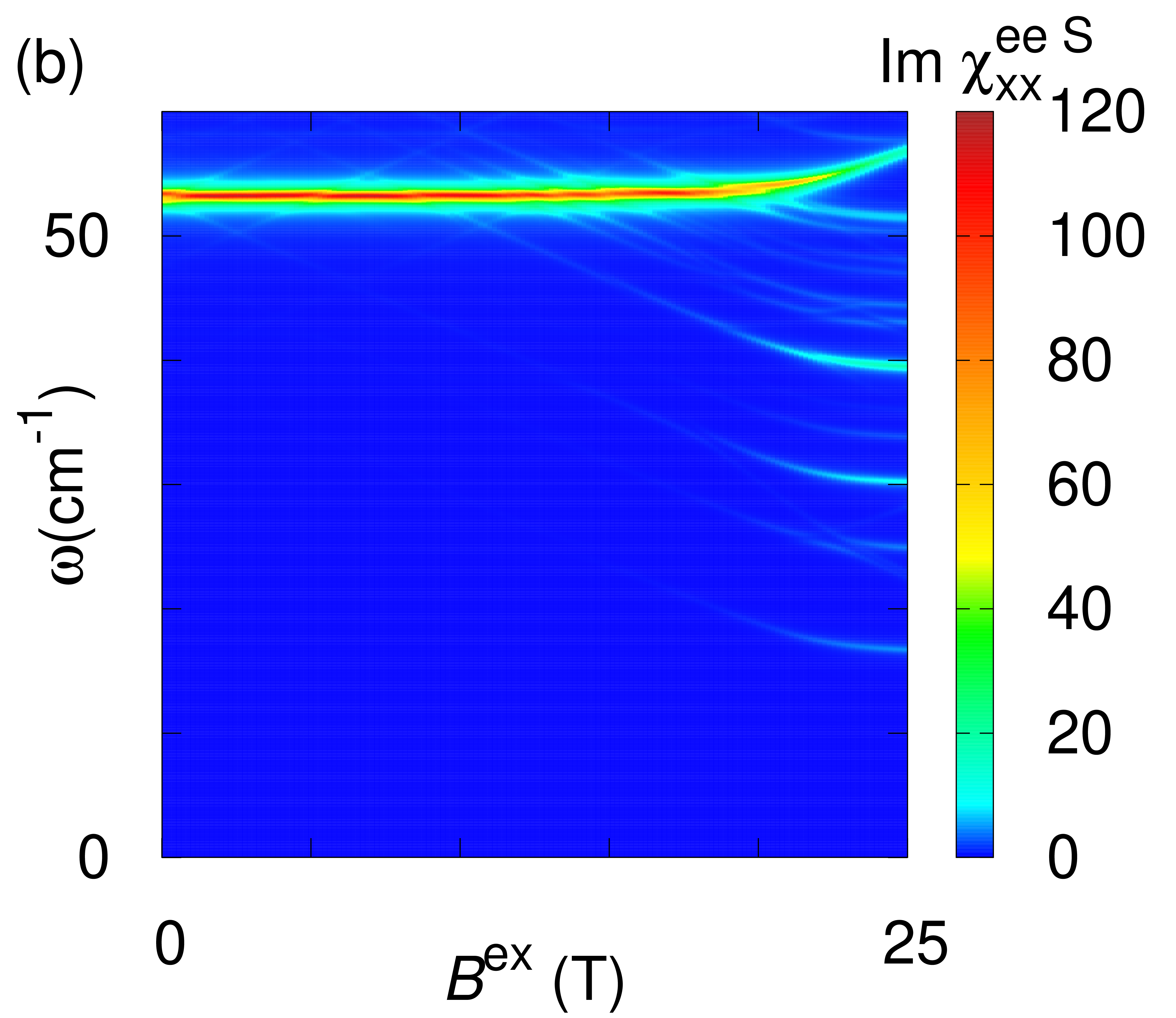} \\
      \includegraphics[width=0.48\columnwidth, bb= 0 0 4541 3929]{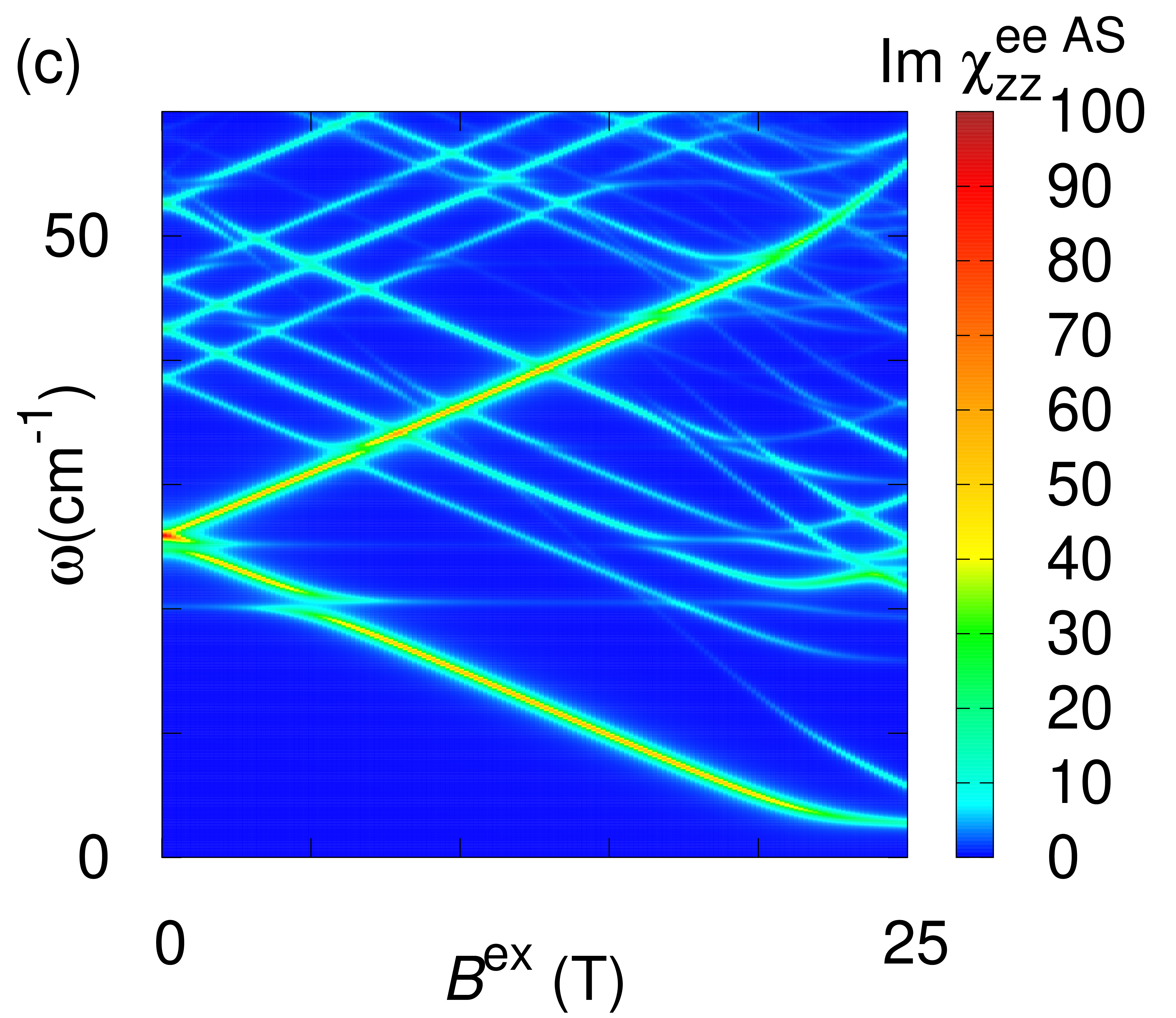}
      \includegraphics[width=0.48\columnwidth, bb= 0 0 4417 3929]{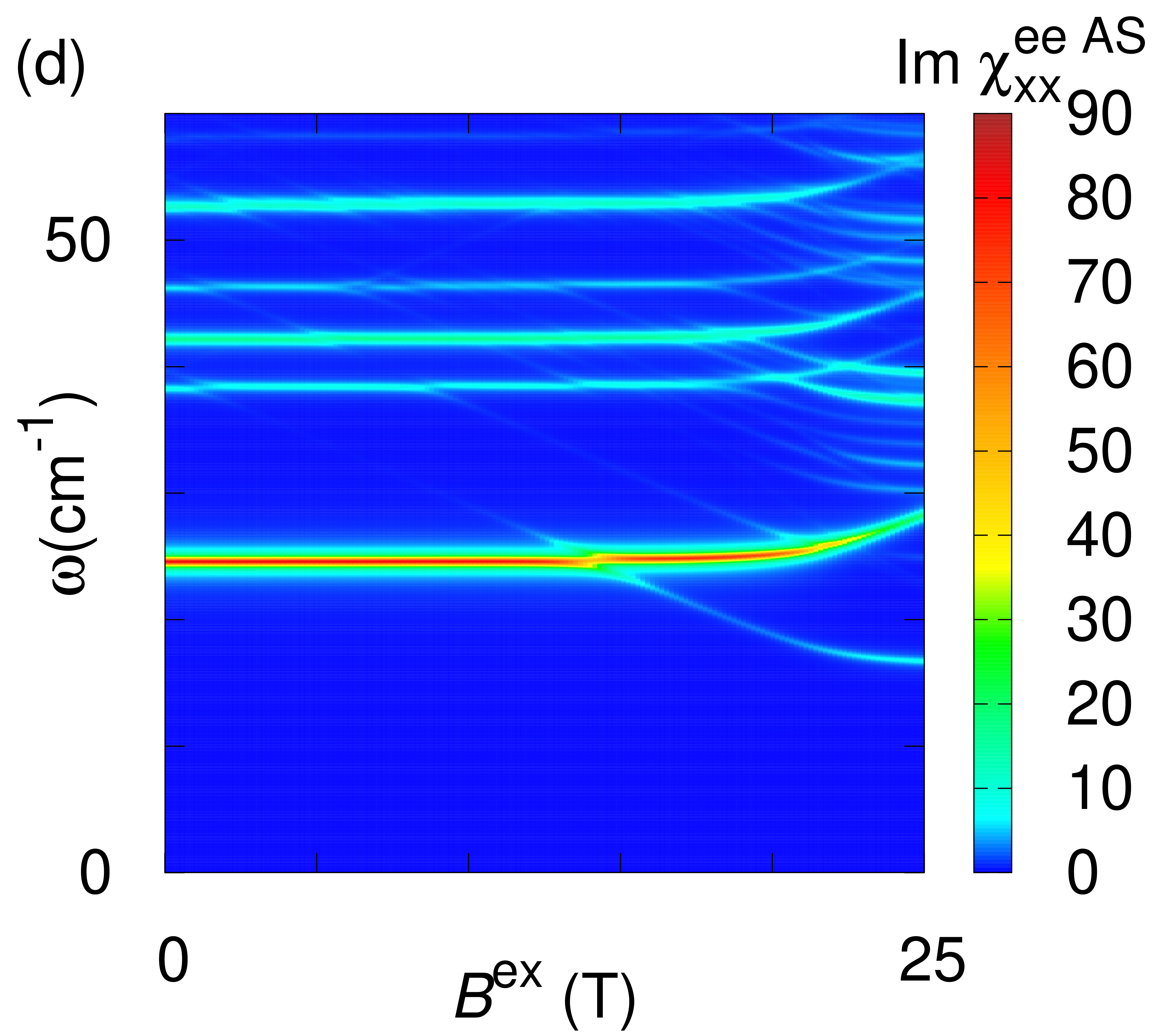} \\
      \includegraphics[width=0.48\columnwidth, bb= 0 0 4541 3929]{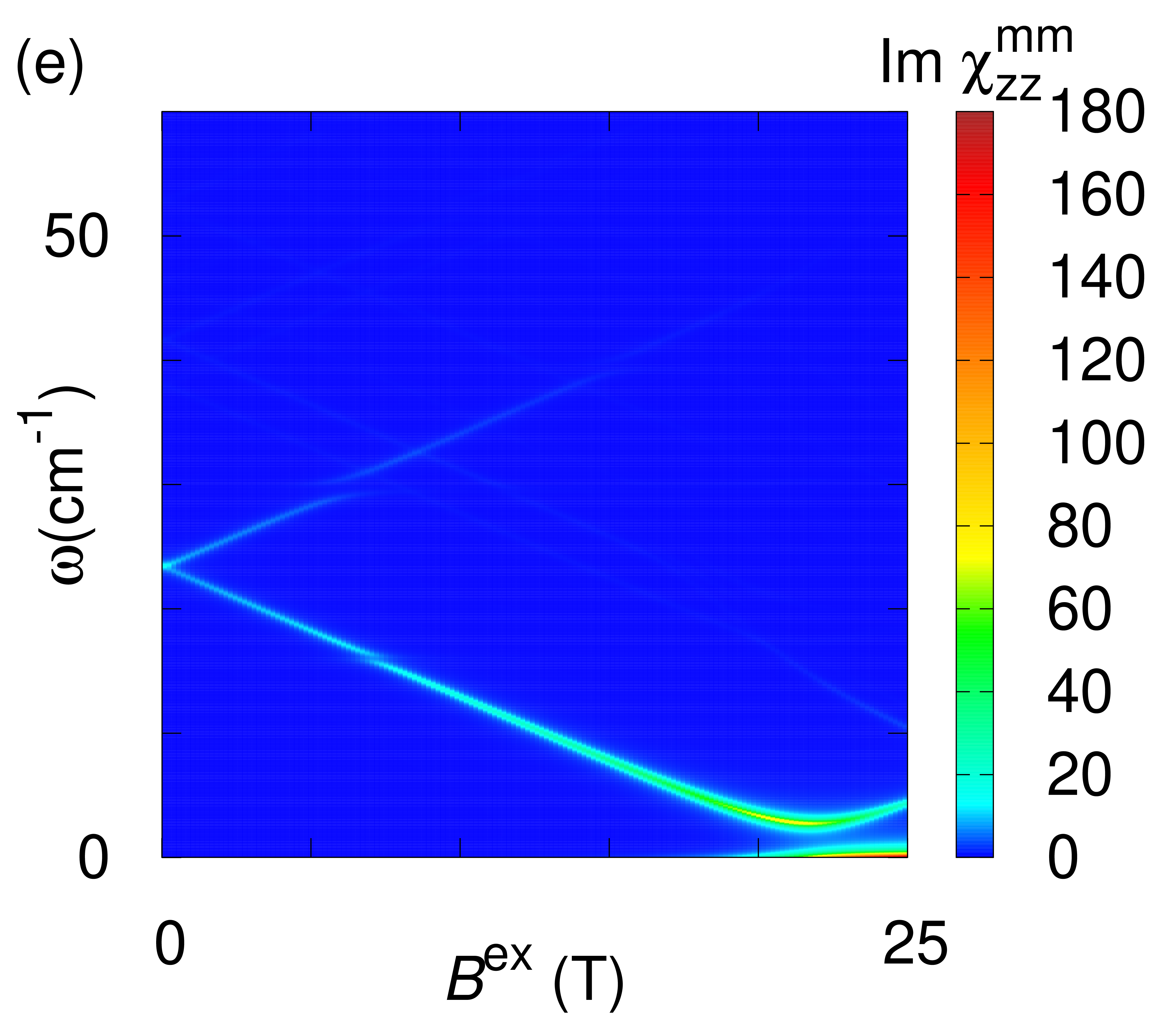}
      \includegraphics[width=0.48\columnwidth, bb= 0 0 4541 3929]{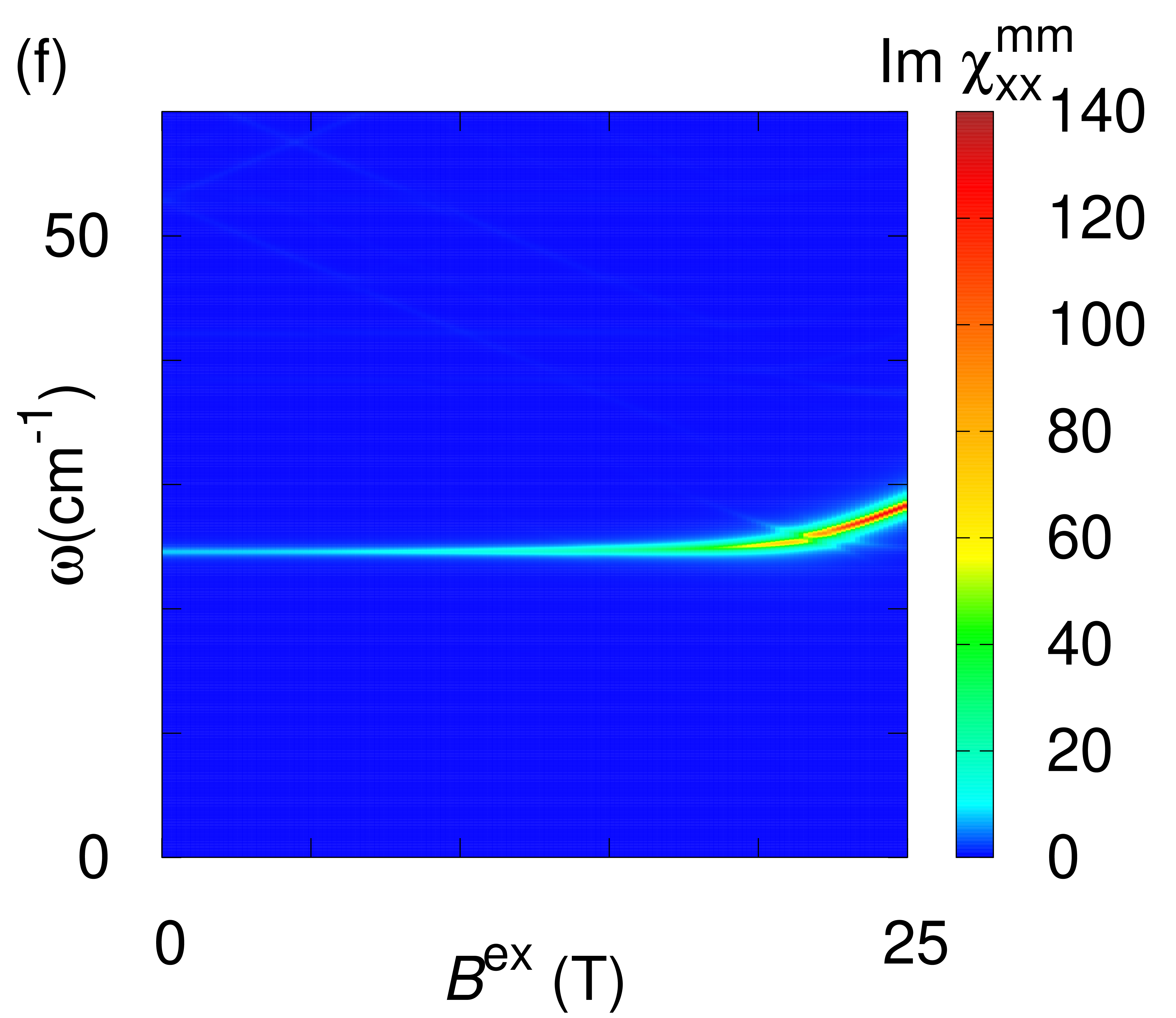}
    \end{center}
    \caption{(Color online)
      The imaginary parts of dynamical susceptibility under an external magnetic field ${\bm B}^{\rm ex} \| z$ in an $N=20$ cluster. The parameters are fixed to be $J^\prime/J = 0.635$, $D/J = 0.034$, $D^\prime/J = 0.02$, and $J = 85$ K.
      (a) The peak positions of the imaginary parts of dynamical electric susceptibility $\chi^{{\rm ee} {\rm S}}_{xx} (\omega)$, $\chi^{{\rm ee} {\rm AS}}_{xx} (\omega)$, and $\chi^{{\rm ee} {\rm AS}}_{zz} (\omega)$, and dynamical magnetic susceptibility $\chi^{\rm mm}_{x x} (\omega)$ and $\chi^{\rm mm}_{z z} (\omega)$.
      (b) Im $\chi^{{\rm ee} {\rm S}}_{xx} (\omega)$.
      (c) Im $\chi^{{\rm ee} {\rm AS}}_{xx} (\omega)$.
      (d) Im $\chi^{{\rm ee} {\rm AS}}_{zz} (\omega)$.
      (e) Im $\chi^{\rm mm}_{x x} (\omega)$.
      (f) Im $\chi^{\rm mm}_{x x} (\omega)$.}
    \label{fig:Bz_N20}
  \end{figure}

\begin{figure}
    \begin{center}
      \includegraphics[width=0.48\columnwidth, bb= 0 0 4209 3531]{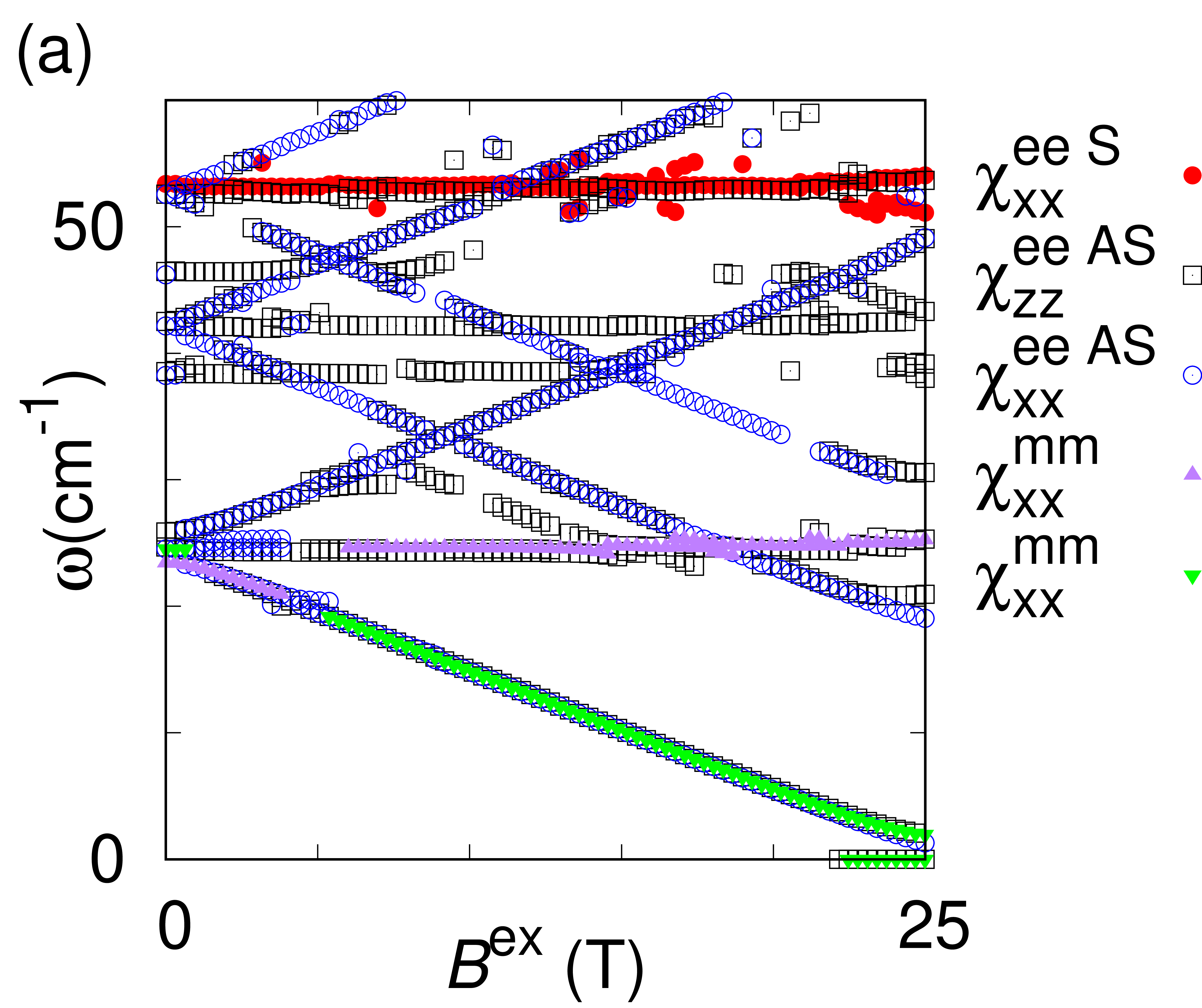}
      \includegraphics[width=0.48\columnwidth, bb= 0 0 4541 3929]{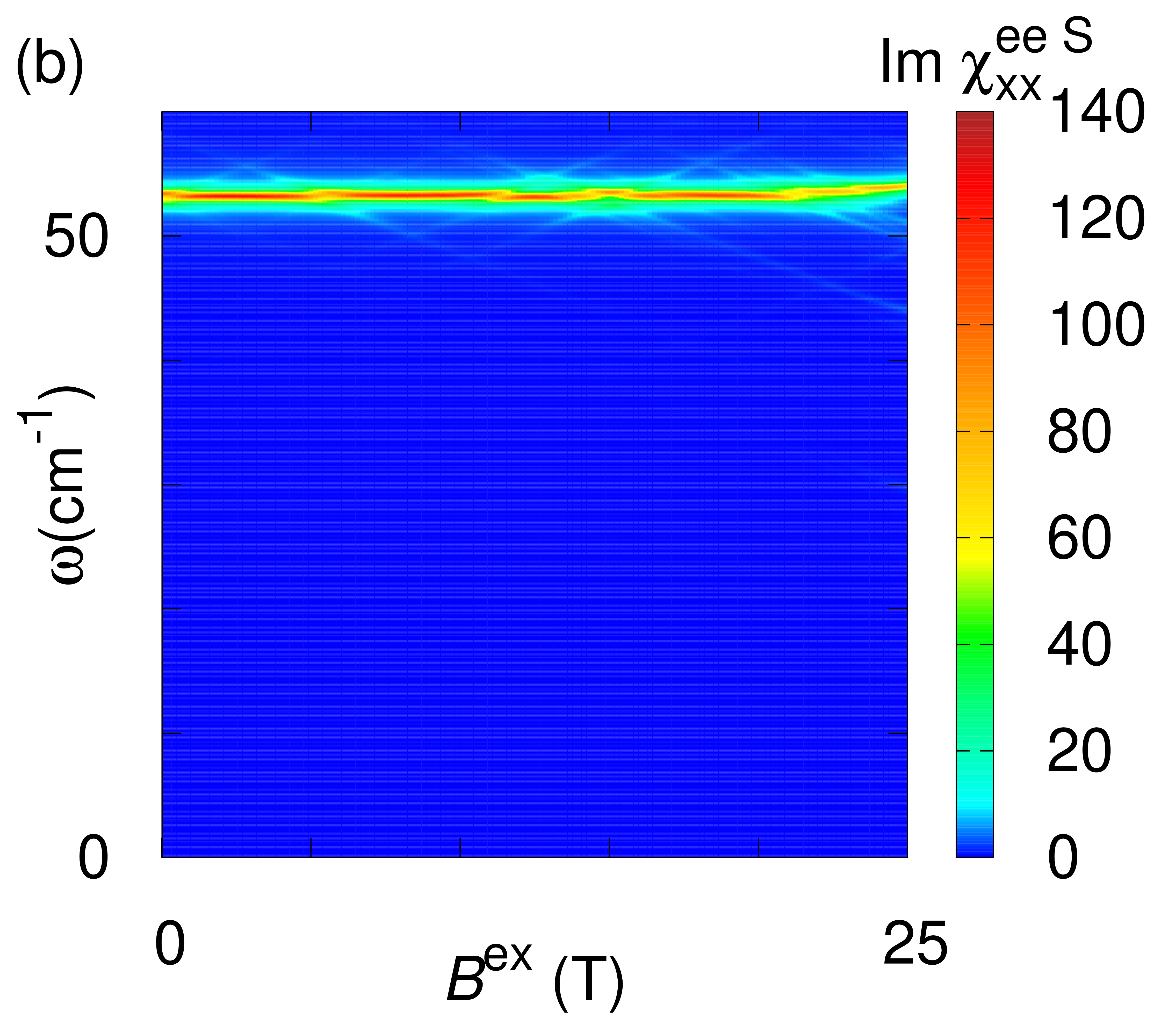} \\
      \includegraphics[width=0.48\columnwidth, bb= 0 0 4541 3929]{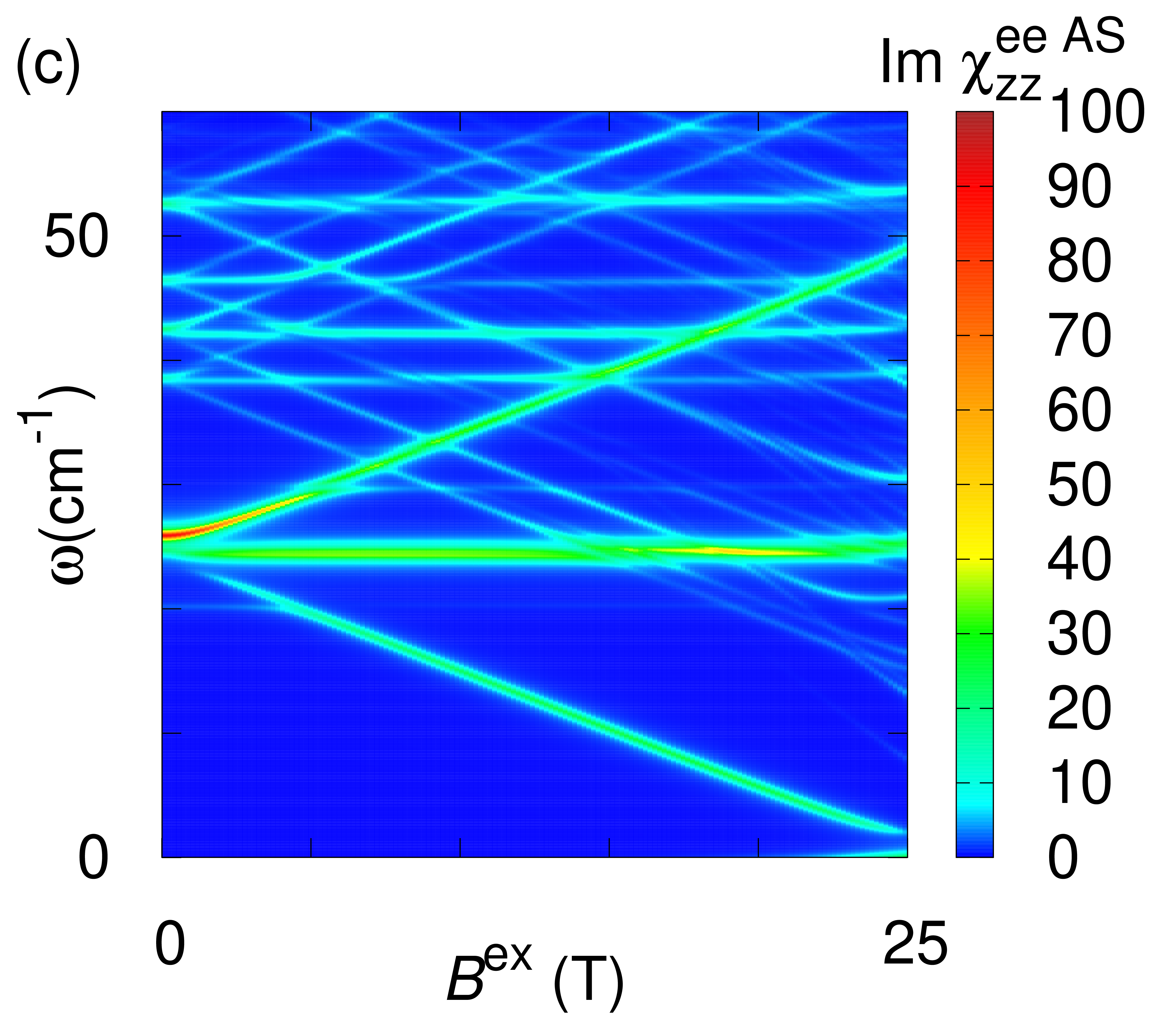}
      \includegraphics[width=0.48\columnwidth, bb= 0 0 4417 3929]{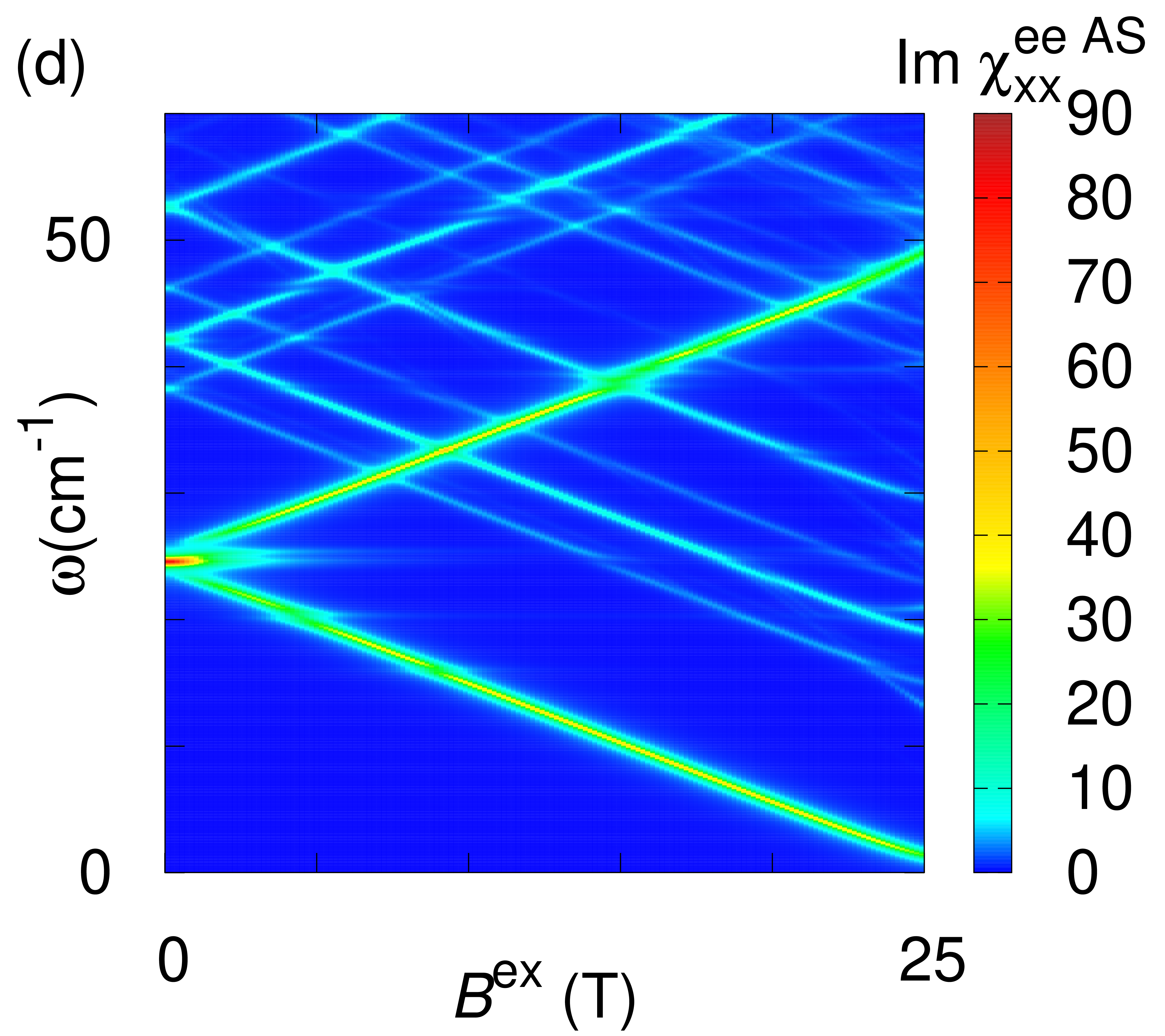} \\
      \includegraphics[width=0.48\columnwidth, bb= 0 0 4411 3929]{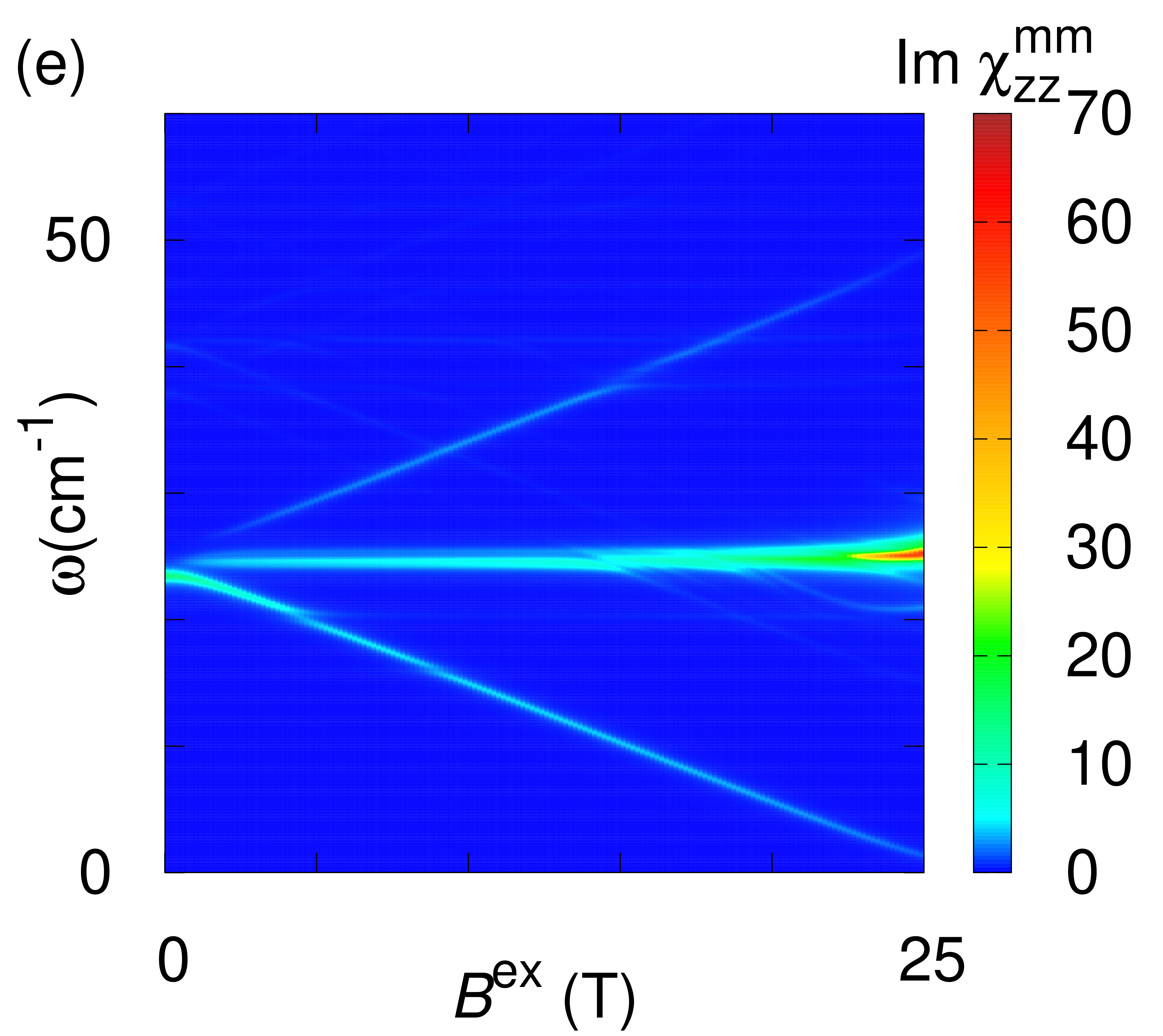}
      \includegraphics[width=0.48\columnwidth, bb= 0 0 4411 3929]{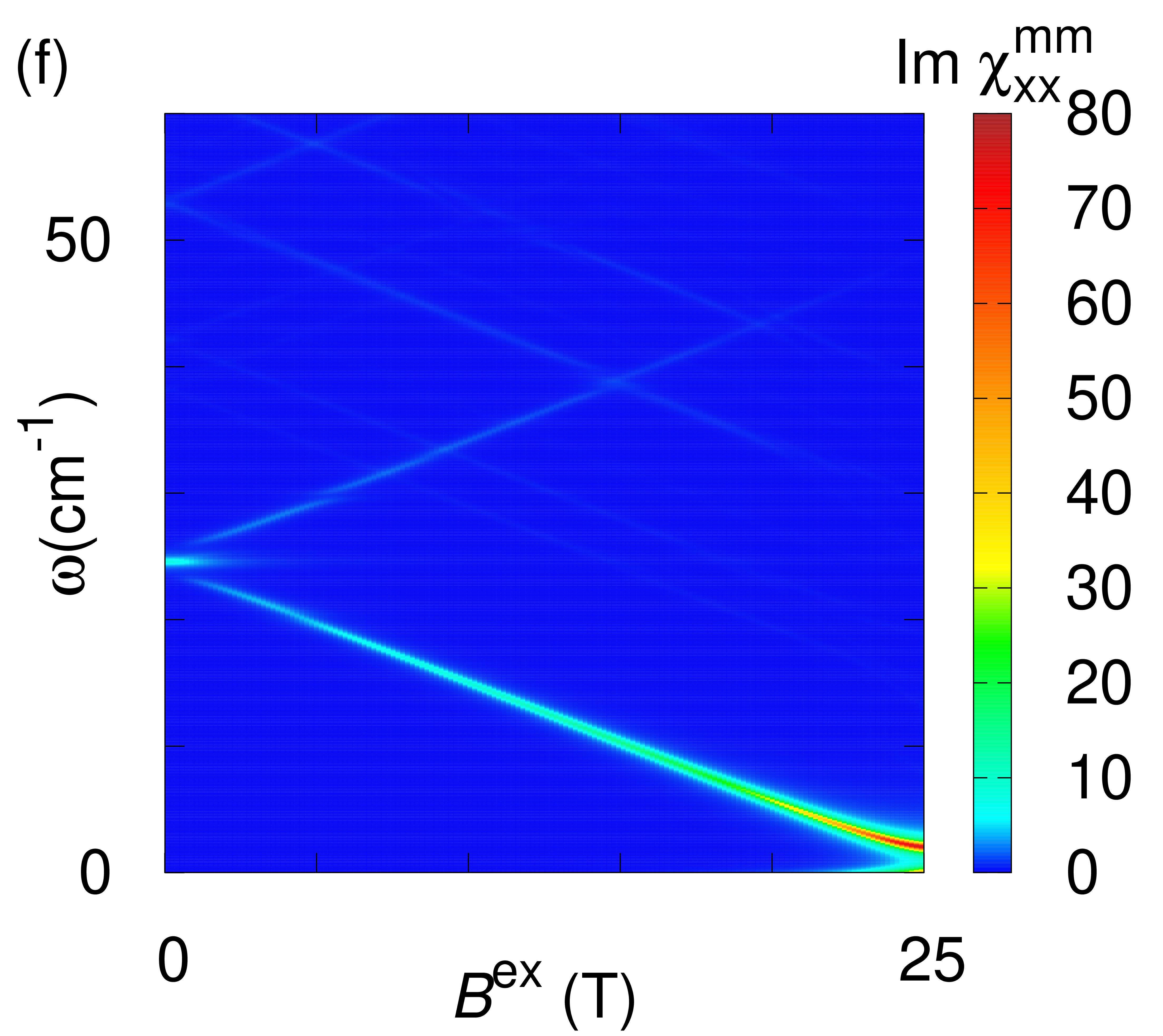} 
    \end{center}

    \caption{(Color online)
      The imaginary parts of the dynamical susceptibility under an external magnetic field ${\bm B}^{\rm ex} \| x$ in the $N=20$ cluster. The parameters are fixed to be $J^\prime/J = 0.635$, $D/J = 0.034$, $D^\prime/J = 0.02$, and $J = 85$ K.
      (a) Peak positions of imaginary parts of the dynamical electric susceptibility $\epsilon^{\rm S}_{xx} (\omega)$, $\epsilon^{\rm AS}_{xx} (\omega)$, and $\epsilon^{\rm AS}_{zz} (\omega)$ 
     , and dynamical magnetic susceptibility $\mu_{x x} (\omega)$      and $\mu_{z z} (\omega)$.
      (b) Im $\epsilon^{\rm S}_{xx} (\omega)$.
      (c) Im $\epsilon^{\rm AS}_{xx} (\omega)$.
      (d) Im $\epsilon^{\rm AS}_{zz} (\omega)$.
      (e) Im $\mu_{x x} (\omega)$.
      (f) Im $\mu_{x x} (\omega)$.}
    \label{fig:Bx_N20}
  \end{figure}

  For further comparison with ESR and FIR experiments, dynamical susceptibility under external magnetic fields $B^{\rm ex} \| x$ and $z$ was calculated using the Lanczos method in the $N=20$ cluster. Figures~\ref{fig:Bz_N20} and \ref{fig:Bx_N20} show the peak positions and spectra of the imaginary parts of dynamical susceptibilities. When ${\bm B}^{\rm ex} \| z$, the triplon modes split into three, higher energy $S_z = \pm 1$, lower energy $S_z = \pm 1$, and almost degenerate $S_z = 0$ modes, due to interdimer DM interaction $D^{\prime}$\cite{cepas01}.
  The results indicate that higher energy $S_z = \pm 1$ modes are electroactive, and the lower energy $S_z = \pm 1$ modes are magnetoactive. $S_z = 0$ modes are electroactive and magnetoactive. When ${\bm B}^{\rm ex} \| x$, the triplon modes split into three, $S_x = 1$, $S_x = 0$, and $S_x = -1$ modes\cite{cepas01}, and all modes are electroactive and magnetoactive. The results indicate that the $S^{\rm tot}=1$ bound states of two triplons (triplet triplon pairs) are excited primarily through electroactive processes. Since the triplet triplon pair can move due to correlated hoppings~\cite{momoi00,momoi00b,fukumoto00b,totsuka01}, the system size effects cannot be neglected, and the quantitative comparison of the peak positions between the experimental and theoretical results is challenging. However, the qualitative features are consistent.
  At around the critical field, resonances due to the $S^{\rm tot}=2$ bound states of two triplons (quintet triplon pairs) are active due to the electric components [Fig.~\ref{fig:Bz_N20}(c)], consistent with the ESR observations~\cite{nojiri03}. Therefore, the quintet pair can be the primitive excitation at around the $1/8$-plateau~\cite{corboz14}.
 
  In summary,  we achieved a comprehensive understanding of the magnetic excitations observed by ESR and FIR in ${\rm SrCu_2(BO_3)_2}$ by considering the effects of the magnetoelectric couplings and the DM interactions in the Shastry-Sutherland model. The dominant absorption processes in ${\rm SrCu_2(BO_3)_2}$ are electro-triplon at approximately 25 ${\rm cm^{-1}}$ and the $S^{\rm tot} = 0$ bound states of two triplons at approximately 55 ${\rm cm^{-1}}$. The latter process can be active only through electromagnetic couplings. In this way, clarifying the electroactive magnetic excitation in various nonmagnetic spin states will be crucial to find novel magnetic excitations from a theoretical viewpoint. Furthermore, establishing the theory of electroactive magnetic excitations in a nonmagnetic spin state can lead to observing novel magnetic excitations in experiments.
  
  The transitions between the spin-singlet ground state and the triplet excitations were observed in various spin gap systems, {\it e.g.}, the spin-Peierls material ${\rm CuGeO_3}$~\cite{brill94,nojiri99b}, the spin ladder ${\rm Sr_{14}Cu_{24}O_{41}}$~\cite{huvonen07}, the Haldane material ${\rm Ni(C_3H_{10}N_2)_2NO_2ClO_4}$~\cite{hagiwara96}, and the trellis lattice ${\rm NaV_2O_5}$~\cite{luther98}. So far, the theoretical analysis of such a forbidden transition in the Heisenberg model is based on the magnetoactive process in a spin model with anisotropic interaction terms~\cite{sakai00,oshikawa03}. This Letter indicates that these singlet-triplet transitions can be understood naturally through electroactivity, even in a Heisenberg model.
  
  Since the electroactive process can exist in anomalous nonmagnetic spin states, {\it e.g.}, quantum spin liquid and spin nematics, there is room to excite novel magnetic excitations, which cannot be excited by magnetic processes and neutron scattering experiments. Confirming the excitation modes and selection rules in various magnets will be crucial for searching the novel magnetic excitations. Moreover, clarifying the electroactive processes in quantum magnets will help control the quantum spin states, {\it i.e.}, in condensed matter physics and quantum information science.
 
\begin{acknowledgments}
The authors thank Nobuo Furukawa and Isao Maruyama for stimulating discussions. This work was supported by Japan Society for the Promotion of Science (JSPS) KAKENHI Grants No. 22H01171. 
\end{acknowledgments}

\bibliographystyle{apsrev}

\end{document}